\documentclass[12pt,manuscript]{aastex}

\slugcomment{Accepted to ApJ Letters}

\def\kms{\ifmmode{\rm km\thinspace s^{-1}}\else km\thinspace s$^{-1}$\fi}
\def\ms{\ifmmode{\rm m\thinspace s^{-1}}\else m\thinspace s$^{-1}$\fi}

\newcommand{\msun}{\ensuremath{M_\sun}}

\shortauthors{Howell, et al.}
\shorttitle{Imaging of TRAPPIST-1}

\begin{document}

\title{Speckle Imaging Excludes Low-Mass Companions Orbiting the Exoplanet Host Star TRAPPIST-1}

\author{
Steve~B.~Howell\altaffilmark{1} \\
NASA Ames Research Center, Moffett Field, CA 94035, USA \\
Mark E. Everett\altaffilmark{1} \\
National Optical Astronomy Observatory, 950 N. Cherry Ave., Tucson, AZ 85719, USA \\
Elliott P. Horch\altaffilmark{2} \\
Department of Physics, Southern Connecticut State University, 501 Crescent St., New Haven, CT, 06515, USA \\ 
Jennifer G. Winters\altaffilmark{1} \\
Harvard-Smithsonian Center for Astrophysics, Cambridge, MA, 02138, USA \\
Lea Hirsch\altaffilmark{1} \\
Astronomy Department, University of California, Berkeley, 510 Campbell Hall, Berkeley, CA, 94720, USA \\
Dan Nusdeo\altaffilmark{1} \\ 
Department of Physics and Astronomy, Georgia State University, P.O. Box 5060, Atlanta, GA 30302, USA\\
Nicholas J. Scott\altaffilmark{1} \\
NASA Ames Research Center, Moffett Field, CA 94035, USA \\
}

\altaffiltext{1}{Visiting Astronomer, Gemini-South Observatory}
\altaffiltext{2}{Adjunct Astronomer, Lowell Observatory}

\begin{abstract}
We have obtained the highest resolution images available of TRAPPIST-1 using the Gemini-South telescope and our 
speckle imaging camera. Observing at 692 and 883 nm, we reached the diffraction limit of the telescope providing a best resolution of 27 mas or, at the distance of TRAPPIST-1, a spatial resolution of 0.32 AU. Our imaging of the star extends from 0.32 to 14.5 AU. We show that to a high confidence level, we can exclude all possible stellar and brown dwarf companions, indicating that TRAPPIST-1 is a single star.
\end{abstract}


\section{Introduction} 

Recently, a near-by, ultra-cool dwarf was observed and discovered to host at least three transiting Earth-sized planets in short orbital periods. The discovery of planets transiting TRAPPIST-1 (2MASS J23062928-0502285) by Gillon et al. (2016) has sent exoplanet scientists into a frenzy of follow-up studies and new theoretical investigations. The closeness of TRAPPIST-1 (12 pc; Costa et al., 2006) and its infrared brightness ($K$=10) are very exciting to observers, allowing the opportunity to fully study the M8V host star as well as the planets. Indeed, de Wit et al. (2016) have already performed the first investigation of atmospheric characterization invoking a  favorable alignment of the two inner planets,  using the Hubble Space Telescope to perform transmission  spectroscopy.

Given the small distance to TRAPPIST-1, the star itself has a relatively large proper motion and as such the current sky position of the dwarf star can be searched in detail in archival images to find or rule out any possible confounding background  sources possibly producing or affecting the transits. Similarly, a widely-separated, common proper motion star can be searched for and detected as well, that is a star seen to move in step with TRAPPIST-1. No such background or common motion stars have been found (Gillon et al. 2016).    

As with all exoplanet host stars, high-resolution imaging is one of the key steps in validation, ruling out spatially close, usually bound fainter companions. These companions are difficult to impossible to find through spectroscopic means as they have orbital periods of hundreds to thousands of days and are often too faint to be revealed in a (single) spectrum (see Teske et al., 2015). Our team has been investigating the nature of exoplanet host stars in part to validate their planets as well as to study the stars in their own right to understand the binarity of exoplanet hosts (Howell et al., 2011; Horch et al., 2012, 2014; Ciardi et al., 2015; Furlan et al., 2016)
We have especially targeted stars hosting small, rocky planets essentially impossible to observationally validate in any other manner.
 
We present herein the first high-resolution speckle imaging of TRAPPIST-1 obtained in order to discover or rule out a close, bound stellar companion hiding within the system.  Given the range of orbital periods found for the planets (1.5 to $<$100 days),
a companion star would need to orbit farther out from the M8V host star to ensure the stability of the planetary system. An as yet unknown bound star would be lower in luminosity 
and have an angular separation and relative magnitude rendering it undetected in
the reported TRAPPIST-1 observations.

\section{Observations}

We observed TRAPPIST-1 using our visiting Differential Speckle Survey Instrument (DSSI) speckle camera mounted on the 8-m Gemini-South Telescope. TRAPPIST-1 is a M8V star with $V$=18.8, $R$=16.5, $I$=14.0 and $J$=11.4 (Costa et al., 2006). DSSI  
is a dual-channel, fast readout imager providing a 2.8 arcsec field of view and 
simultaneous images in two different bands. DSSI as used at the Gemini-North telescope
is discussed in Horch et al. (2012)  and our observations at Gemini-South proceeded in a similar manner.

TRAPPIST-1 was observed on the nights of 22 and 27 June 2016 in three medium band filters with central 
wavelengths and bandpass FWHM values of ($\lambda_c,\delta\lambda$)=(562, 43), (692,47) and (883,54) nm.
The first night of observations used a total of 30 minutes of telescope time and consisted of 
12 image sets consisting of 1000 60ms simultaneous frames at 692 nm and 883 nm.  These observations were made during clear weather at airmass 1.10-1.12, when the native seeing was 0.4 arcsec. Observations on 27 June consisted of 6 such image sets each at 562 nm and 883 nm.
We did not detect the source at 562 nm, had a moderate S/N detection at 692 nm, and a high S/N detection at 883 nm, consistent with the very red spectral energy distribution of TRAPPIST-1.
The 883 nm results obtained on 27 June are consistent with those we obtained 5 nights earlier but the 22 June observation provides a deeper image with higher S/N.

Figure 1 shows our reconstructed 692 nm and 883 nm high-resolution speckle images of TRAPPIST-1, showing that it is a single source. We present both linear and max/min log scaled versions of the images. 
Given the moderate S/N obtained in the 692 nm image, we used a low-pass filter that decreases the final resolution of this image ($\sim$40 mas) compared with the full resolution at the diffraction limit (27 mas) of the 883 nm image. 

Figure 2 presents our quantitative imaging results, providing the 5$\sigma$ delta magnitude lower limit we can set for any undetected companion stars from 0.027 arcsec radially outward to 1.2 arcsec. The structure of these quantitative plots are discussed in Howell et al. (2011) and we note that the occasional square points near the 5$\sigma$ line are all associated with the faint horizontal line artifact remaining in the reconstructed images and are not faint stars (see the linear stretch images in Fig. 1).
At the distance of TRAPPIST-1 these angular dimensions convert to the radial dimensions of 0.32 to 14.5 AU and correspond to approximate orbital periods of $\sim$215 days to $\sim$200 years, assuming a host star mass of 0.08$\msun$.  At farther distances, $>$1.2 arcsec, a ground-based image with 0.5 arcsec or so seeing could detect most companions. Thus, our images set strict limits on the lack of bound companions interior to regions of the sky amenable to seeing-limited images and outside typical casual and existing RV detection.
 
\section{Discussion}

We know from Gillon et al. (2016) based on the high proper motion of  TRAPPIST-1, that no unbound, background star exists 
at the current sky position that might be the cause of the transit-like events detected, and the multi-planet nature of this system provides additional assurance that the planets are real (Rowe et al., 2014). However, there is the possibility of a bound, low mass companion star to TRAPPIST-1. Such a companion of equal or lower mass and luminosity would need to orbit with a period of at least $\sim$3.5 times that of the outermost planet in order to maintain planet stability 	
(Holman and Wiegert, 1999). Note, the presence of such a companion would not invalidate the transit events as being due to planets but would suggest that their actual radii are
larger than that implied by the measured transit depths seen in the light curve (see Ciardi et al., 2015; Hirsch et al., 2016).
Given the outermost  planet's (TRAPPIST-1d) longest possible and most likely orbital periods (72 and 18 days; Gillon et al. 2016), the stability criteria requires that no stellar mass object can be in a circular orbit with a period of less than about 60-250 days or approximately an orbital semi-major axis of 0.1-0.4 AU.

Winters et al. (2015) present both photometric and trigonometric distances for TRAPPIST-1. A comparison of these two measurements can indicate whether the star is over-luminous for its type, that is, if there is an near-equal magnitude companion present. The two reported distances agree: 11.6$\pm$1.8 pc for the photometric distance and 12.1$\pm$0.4 pc for the parallax distance. Therefore, TRAPPIST-1 is not over-luminous for its parallax distance and we can use this fact to say that no similarly bright (delta of $\sim$1 mag) companion exists at any separation. Spectroscopic measurements at $\sim$6 km/sec for TRAPPIST-1 (Reiners \& Basri 2010) also rule out an equal-luminosity companion with a separation smaller than $\sim$1 AU.
In addition, Barnes et al. (2014) performed a radial velocity search for exoplanets of 15 M5-M9 dwarfs, including TRAPPIST-1. While their sensitivity was not able to detect planets, they did set constraints on stellar mass companions at close in separations. These existing RV measurements can be used to rule out stellar mass companions ($\sim$0.03-0.08$\msun$) interior to $\sim$0.1-0.15 AU, near the inner separation probed by our speckle measurements.

Our high resolution 883 nm speckle image of TRAPPIST-1 reveals that no additional stellar flux sources are present near and within the system to a contrast limit of 5.5 magnitudes covering 0.32 AU out to 14.5 AU.
Figure 3 presents our 5$\sigma$ detection likelihood as a function of projected (physical) separation. The curve on the plot is based on the delta magnitude observed (Fig. 2, bottom panel), assumes companions have circular orbits (although eccentricities of even 0.5-0.8 do not make much of a difference), and takes into account any unfortunate timing our observations might have whereby we'd miss being able to resolve a companion star due to its orbital alignment during our observations.
The inner angular separation of TRAPPIST-1 resolved in our image from the 8-m Gemini-South telescope is the diffraction limit of 0.027 arcsec or 0.32 AU. This limit corresponds to an orbital period of $\sim$215 days for any possible low-mass companion. 
The outer limit of 1.2 arcsec (14.5 AU) is distant enough that any companion beyond this limit would be detectable in deep, seeing-limited imaging as a common proper motion star. 

TRAPPIST-1 has absolute magnitudes of M$_V$$\sim$+18.3 and M$_z$$\sim$13.2, the latter value calculated from the data presented in Hawley et al., (2002). Since our 883 nm filter sits in the blue end of the SDSS $z$ band, delta magnitude values observed by us will have approximately the same relative differences as $\Delta$M$_z$ values for known stars and brown dwarfs (Dahn et al., 2002; Galicher et al., 2014). Over half of our imaged separations, we observed no additional flux to $\sim$5.5 magnitudes fainter than TRAPPIST-1 itself, thus, all other main sequence stars, all other L-type brown dwarfs, and all T-type brown dwarfs to about T7 are ruled out down to M$_z$$\sim$18. Inside of $\sim$8 AU, our contrast drops and we can only fully eliminate companions of T6 to L5 respectively as indicated on Figure 3.

Our high-resolution images have shown that over a significant portion of the imaged volume, TRAPPIST-1 is a single host star containing no other bound orbiting stellar mass or L-type brown dwarf mass companions residing within the system from 0.32 to 14.5 AU. Using our detection percentage inside of 8 AU and the photometric and RV constraints mentioned above, a mid to late T-dwarf (T3-8) companion still remains slightly possible. However, even if such a low luminosity source exists, its effect on the planet transit depth (and thus planet radii) will be negligible (see Ciardi et al., 2015).  

\section{Summary}

We have presented the highest resolution 692 nm and 883 nm images available of TRAPPIST-1. Our images cover the spatial range of 27 mas to 1.2 arcsec, corresponding to distances of 0.32 to 14.5 AU at the location of the star. 
The depth of our 883 nm image eliminates essentially all possible companions to TRAPPIST-1.
Taken together with previous photometric and RV results, we can eliminate at high confidence, essentially all close companions orbiting the host star. 
Therefore, we find TRAPPIST-1 to be a single star.

Detailed photometry and a search for additional smaller planets transiting TRAPPIST-1 will be provided by the NASA K2 mission\footnote{http://keplerscience.arc.nasa.gov} (Howell et al., 2014). K2 will observe TRAPPIST-1 in its Campaign 12 observations covering the time period 15 December 2016 to 4 March 2017. The resulting data and light curves will be made public immediately and be of high interest to the exoplanet community. In the near future, GAIA should provide a definitive result to completely rule out stellar companions with orbits up to a few years.

\acknowledgements{
We wish to thank the staff of the Gemini-South Observatory for their kind assistance with our visiting instrument and during our observation run. The referee made a number of suggestions which led to a better presentation.  Elisa Quintana provided helpful discussions on the topic of planet stability under the influence of a disturbing companion star. J. Winters is supported through a grant from the John Templeton Foundation. The opinions expressed here do not necessarily reflect the views of the John Templeton Foundation. These results are based on observations obtained as part of the program GS-2016A-Q-80 at the Gemini Observatory, which is operated by the Association of Universities for Research in Astronomy, Inc., under a cooperative agreement with the NSF on behalf of the Gemini partnership: the National Science Foundation (United States), the National Research Council (Canada), CONICYT (Chile), Ministerio de Ciencia, Tecnolog\'{i}a e Innovaci\'{o}n Productiva (Argentina), and Minist\'{e}rio da Ci\^{e}ncia, Tecnologia e Inova\c{c}\~{a}o (Brazil).
}

{\it Facility:} \facility{Gemini:South}

\begin{figure}
\label{f1}
\includegraphics[angle=0,scale=0.45,keepaspectratio=true]{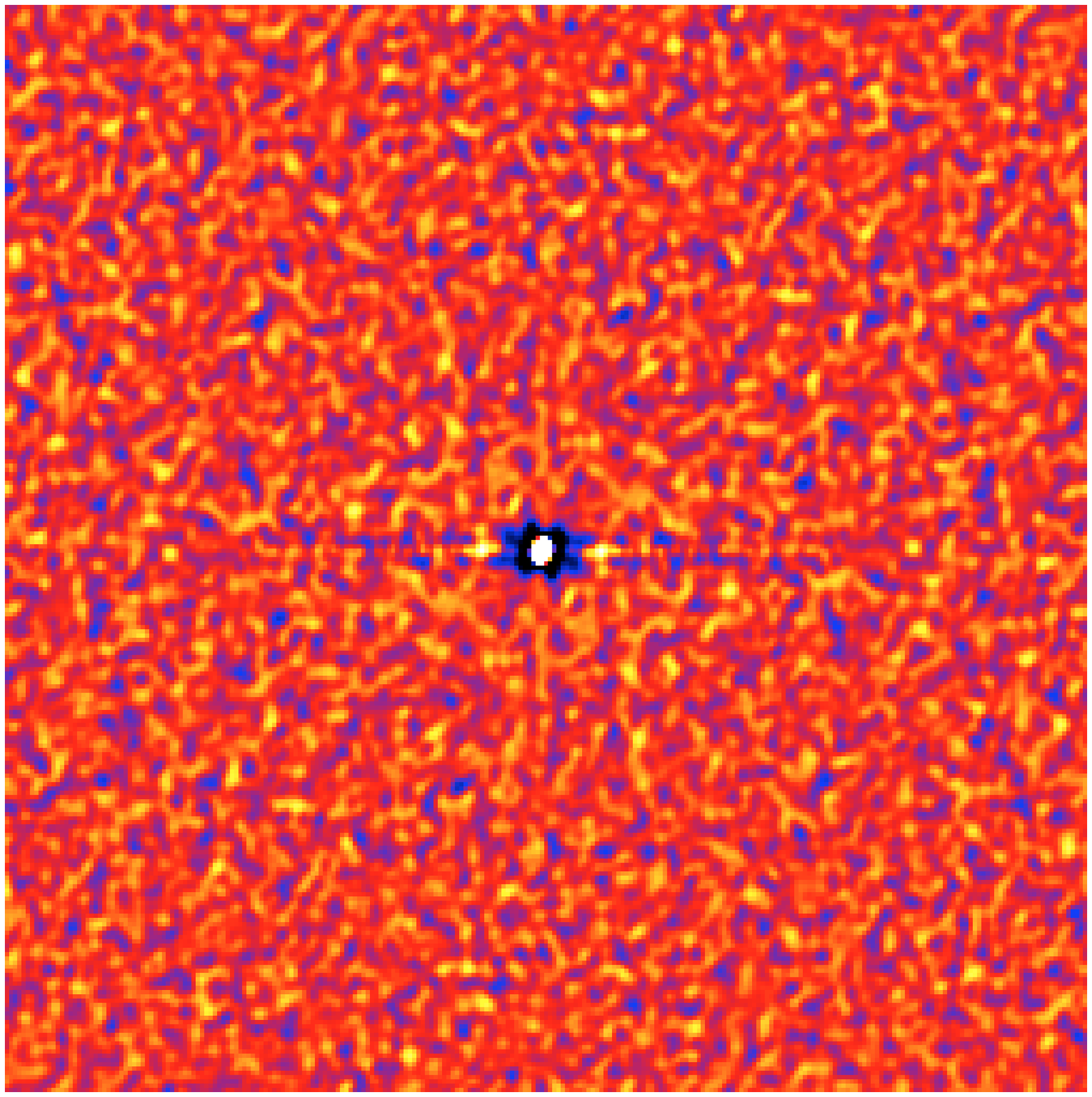}
\includegraphics[angle=0,scale=0.45,keepaspectratio=true]{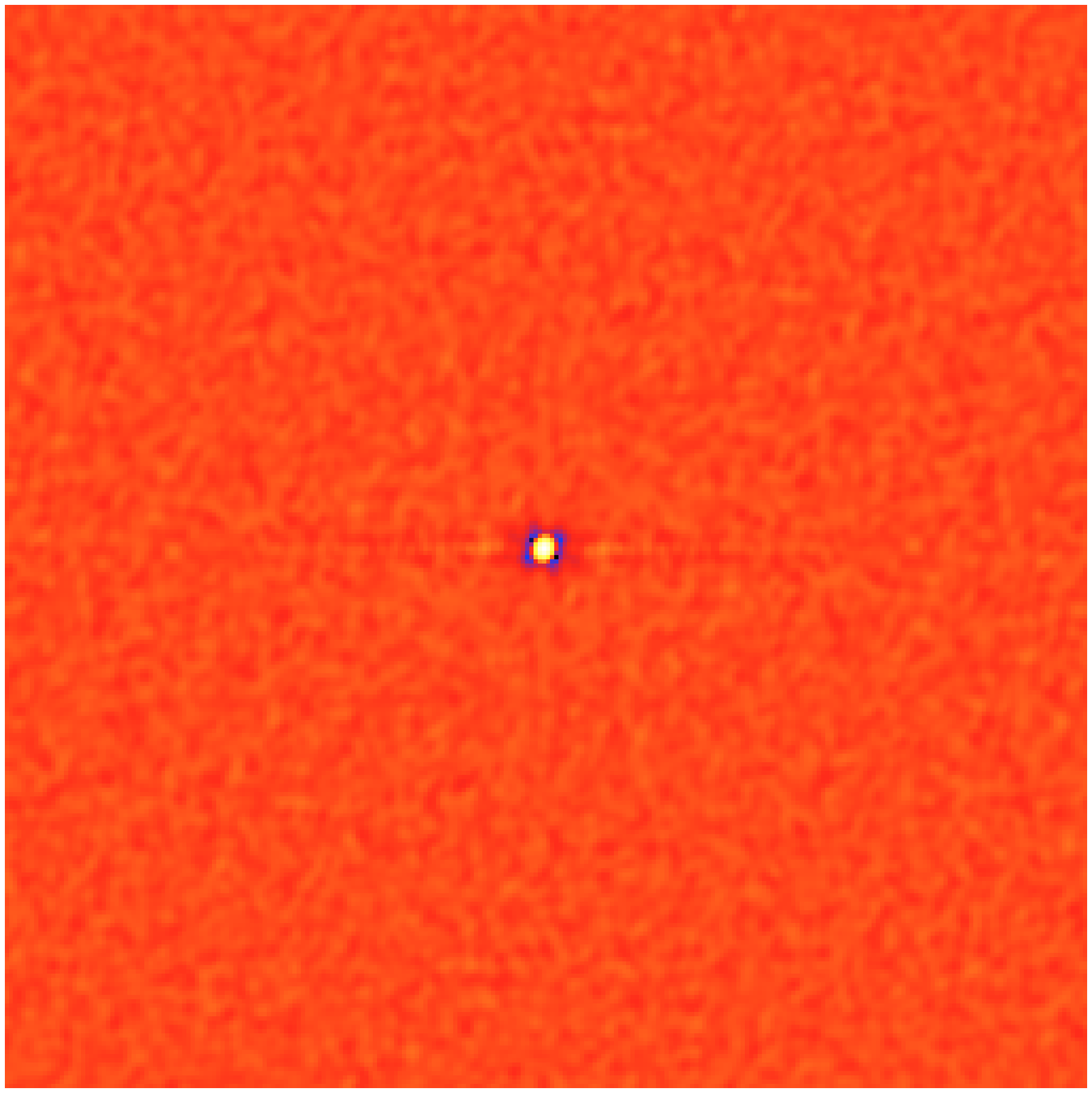}
\includegraphics[angle=0,scale=0.45,keepaspectratio=true]{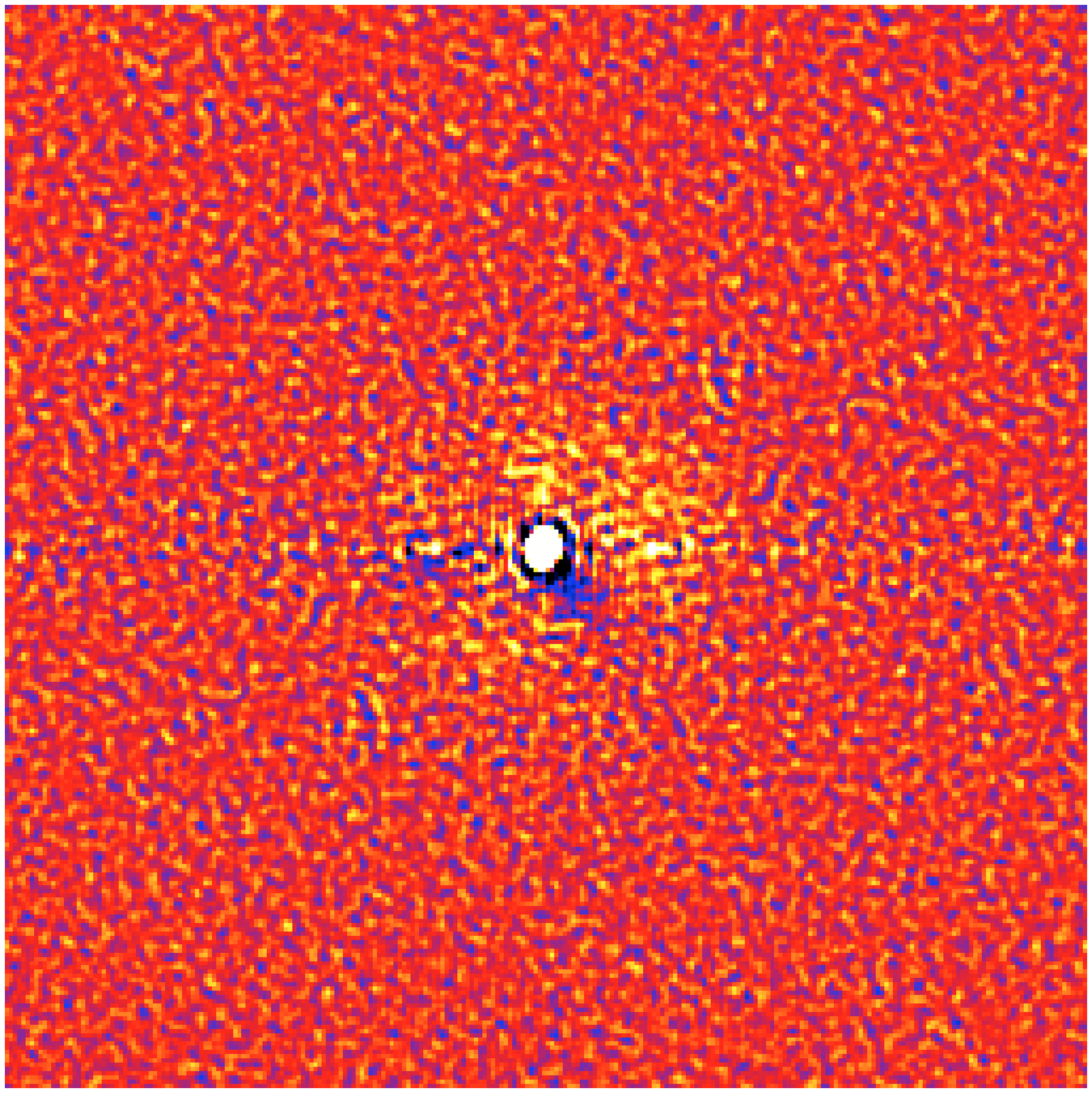}
\includegraphics[angle=0,scale=0.45,keepaspectratio=true]{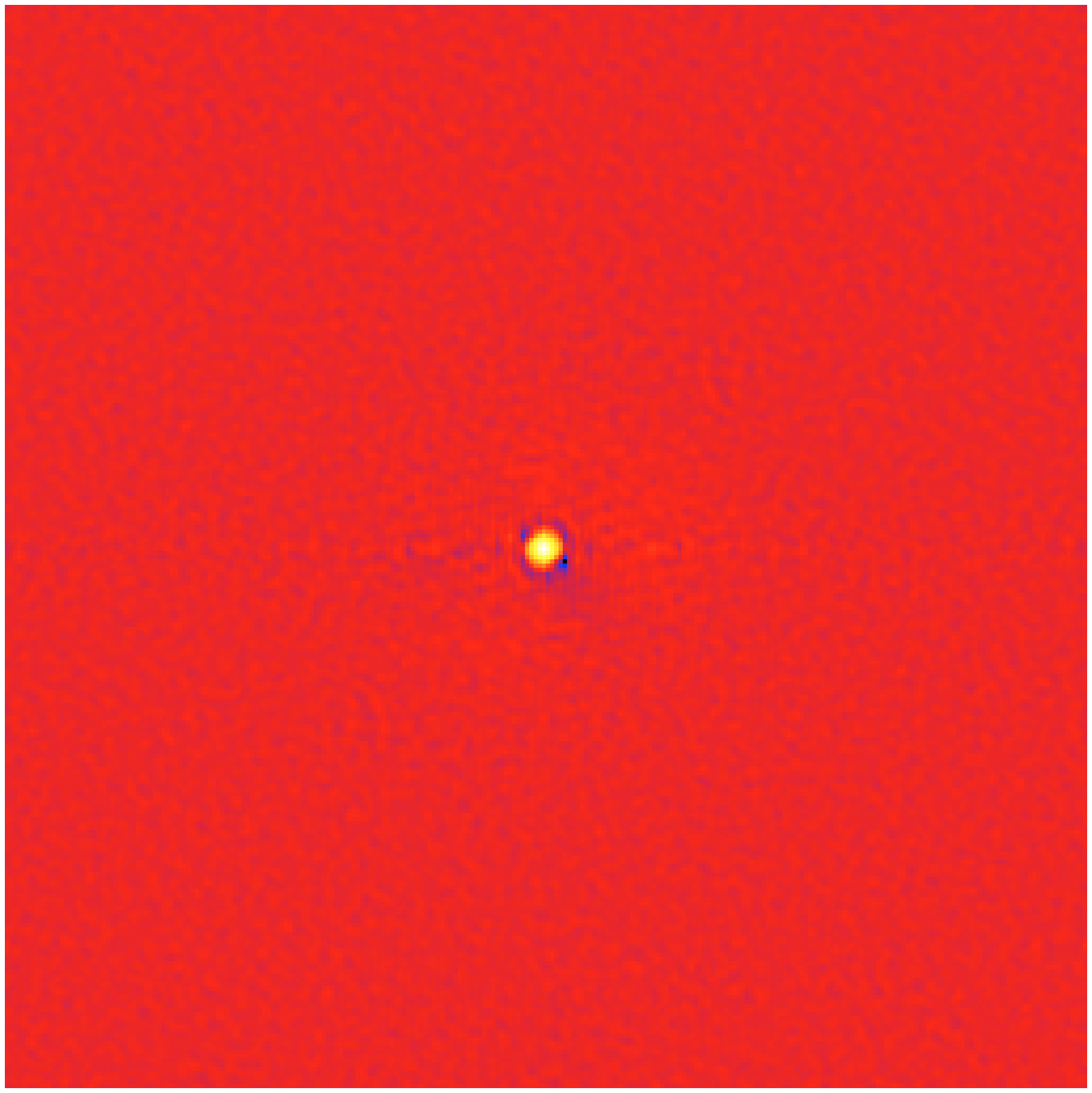}
\caption{The top two panels show the 692 nm reconstructed image obtained on 22 June 2016 of TRAPPIST-1 with both a (left) linear and (right) logarithmic flux scale.
The bottom two panels show the 883 nm reconstructed image obtained on 22 June 2016 of TRAPPIST-1 with both a (left) linear and (right) logarithmic flux scale.
Each image has North up and East to the left and is 2.8 arcsec across (corresponding to a spatial dimension of 34 AU at the distance of TRAPPIST-1). No companion star is detected.}
\end{figure}


\begin{figure}
\label{f3}
\includegraphics[angle=0,scale=0.72,keepaspectratio=true]{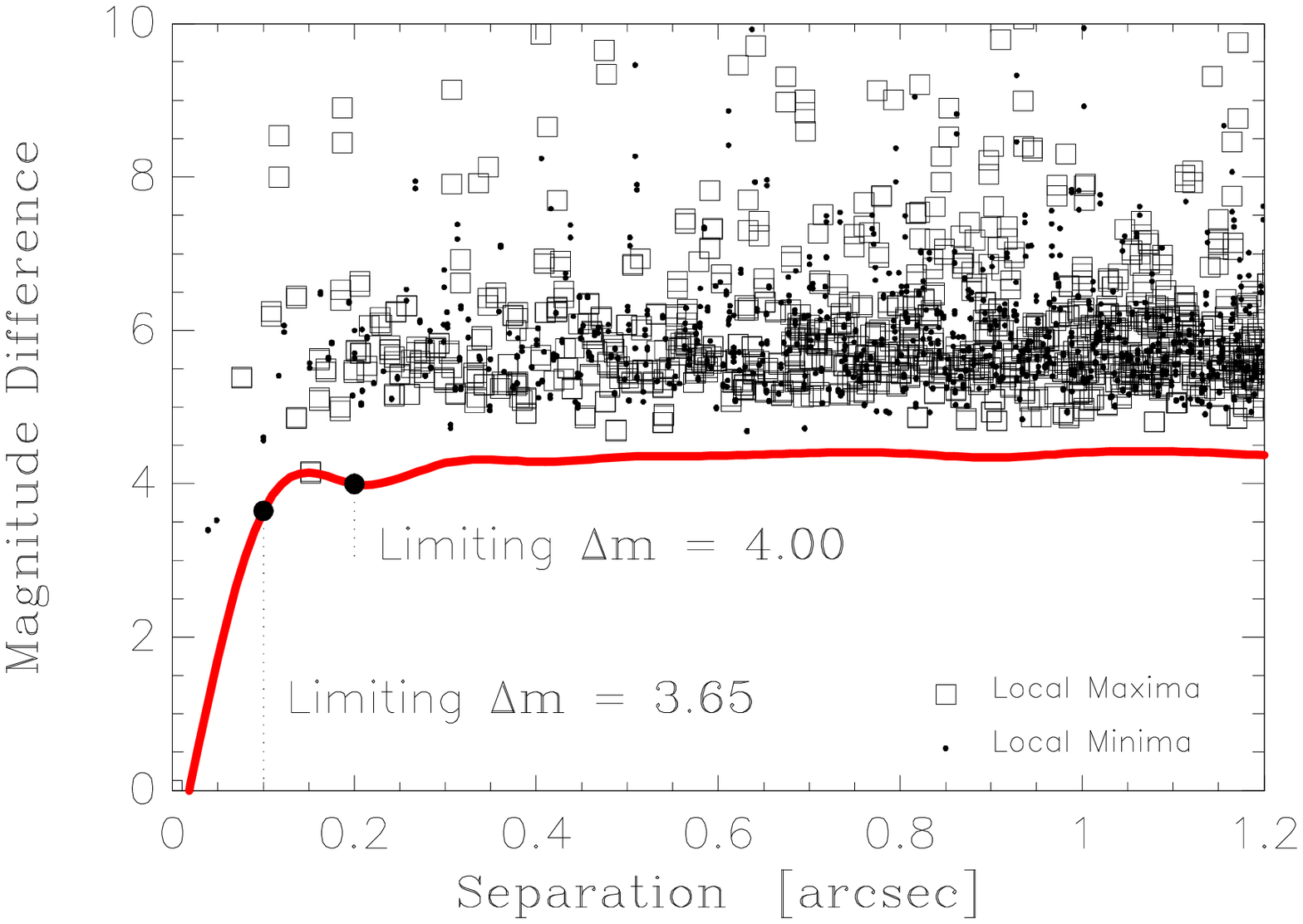}
\includegraphics[angle=0,scale=0.72,keepaspectratio=true]{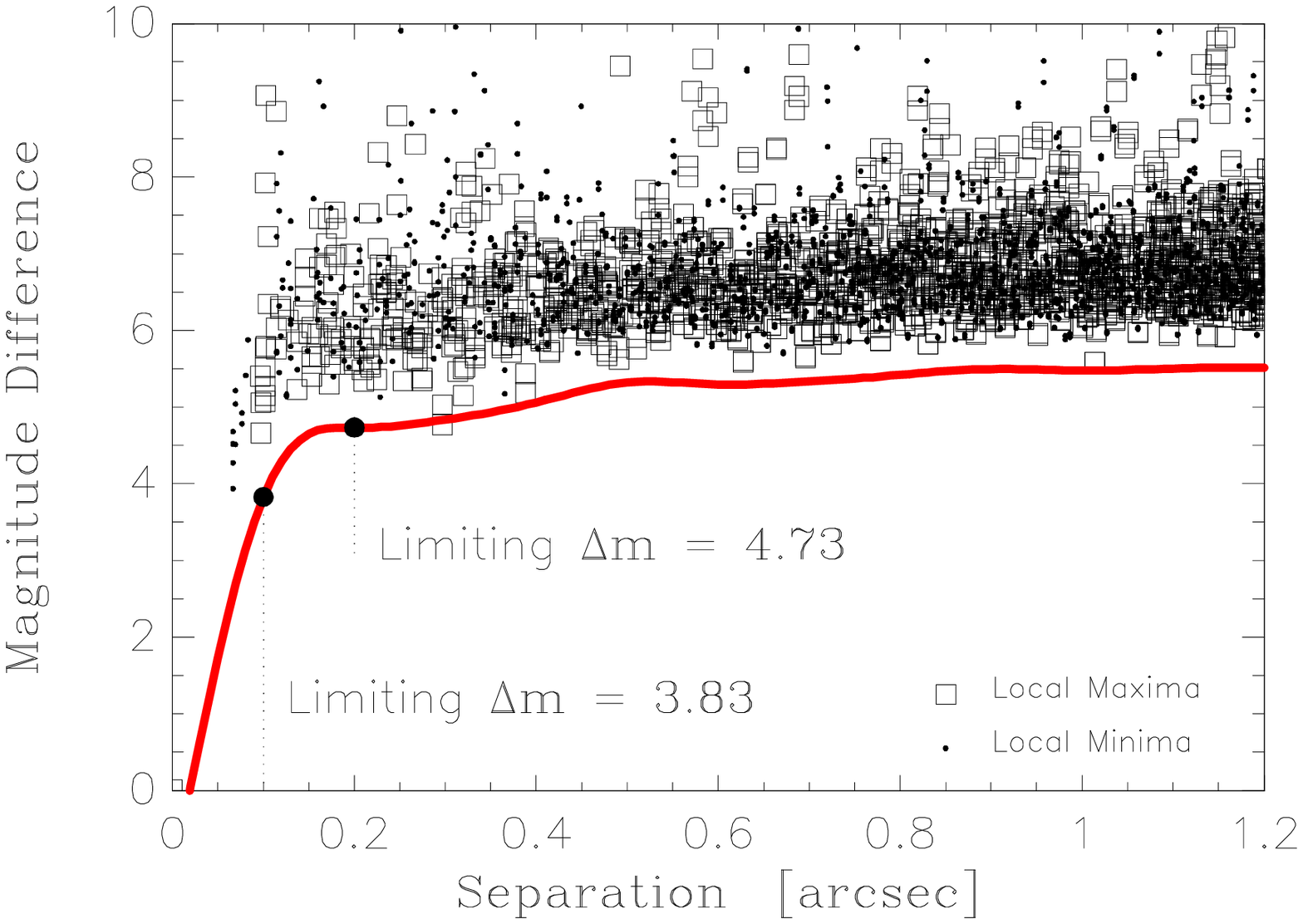}
\caption{Detection limit analysis for the Gemini-South 22 June 2016 observation of TRAPPIST-1.
Detection limits observed at 692 nm (Top) and at 883 nm (Bottom). The red line represents the relative 5$\sigma$ limiting magnitude as a function of separation from 0.027 to 1.2 arcsec. At the distance of TRAPPIST-1 these limits corresponds to 0.32 to 14.5 AU. The two listed limiting magnitudes given for reference are for 
angular separations of 0.1 and 0.2 arcsec.}
\end{figure}

\begin{figure}
\label{f3}
\includegraphics[angle=0,scale=0.72,keepaspectratio=true]{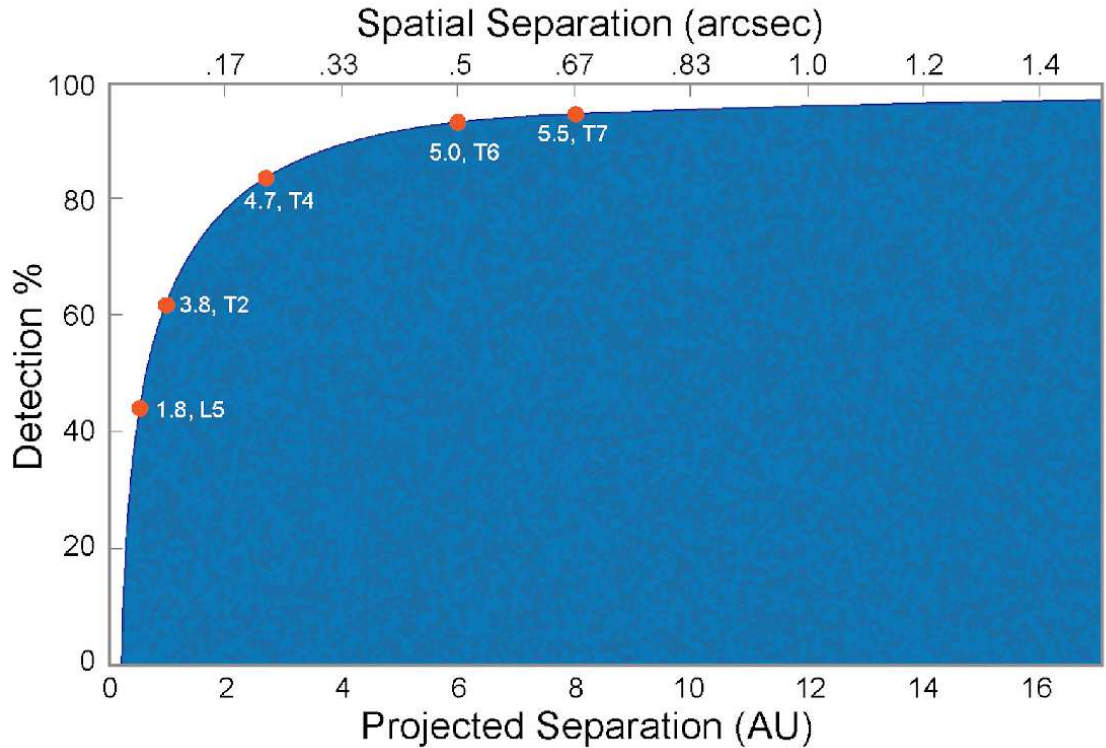}
\caption{Detection percentage as a function of projected angular separation for the 883 nm
speckle image of TRAPPIST-1. The curve is the 5$\sigma$ detection limit from our image
convolved with the detection likelihood (see text). 
The result eliminates all companions in the blue region at separations of 0.32 to 14.5 AU
for the observed contrast ratios. Numbers on the plot represent the delta magnitude and spectral type limits at the corresponding points. For example, all companions earlier than T7 are eliminated to an inner separation of $\sim$8 AU.
}
\end{figure}

\end{document}